\documentclass[a4article,12pt]{article}
\usepackage[T1, T2A]{fontenc}
\usepackage[utf8]{inputenc}
\usepackage[english]{babel}
\usepackage{graphicx}
\usepackage{amssymb,amsfonts,amsmath,mathtext,cite,enumerate,float,microtype,slashed}
\usepackage {indentfirst}
\usepackage {caption}
\usepackage{simpler-wick}
\usepackage{authblk}

\numberwithin{equation}{section}

\title{Loop corrections to the four-fermion interaction}
\author{A.T. Borlakov\thanks{borlakov@theor.jinr.ru} \ and  D.I. Kazakov
\thanks{kazakovd@theor.jinr.ru} }
\affil{Bogoliubov Laboratory of Theoretical Physics, \\ Joint Institute for Nuclear Research, Dubna, Russia}

\date{}

\begin{document}

\maketitle

\abstract{We consider the $f f \rightarrow f f$ scattering amplitudes for the a massless four-fermion interaction model in four dimensions. At first we take the simplest version with the scalar current-current interaction. The loop corrections up to the three-loop level are calculated within the spinor-helicity formalism using the Weyl spinors.  We find out that there are two independent spinor structures that appear in all orders of perturbation theory that can be separated when calculating the diagrams. Our aim is to calculate the leading divergences within the dimensional regularization. To check the validity of our calculations, we use the recurrence relations that connect the leading divergences in the subsequent orders of perturbation theory. We left the derivation of these relations and further analysis of the consequences for another publication for the sake of clarity of the current presentation. At the end we briefly consider the physically more interesting case of V-A interaction.}

\section{Introduction}

The four-fermion interaction is known to be the low energy Lagrangian of weak interactions. It was later replaced by the gauge theory with the intermediate weak bosons for the benefit of renormalizability. Besides, it is well known that at high energy the four-fermion interaction violates unitarity since in perturbation theory it leads to scattering amplitudes increasing with energy. This fact essentially leaves for the four-fermion interaction the role of a low energy effective theory treated basically at the tree level. Without questioning these statements we ask what happens to the amplitudes if one sums up the leading logarithms in all orders of perturbation theory. In doing this we are aware that the four-fermion is non-renormalizable by power counting and the UV divergences, strictly speaking, are not under control. However, as we demonstrated in our earlier papers~\cite{we1,we2}, one can still calculate  radiative corrections and sum up the leading asymptotic in an unambiguous manner. These radiative corrections summed over all orders of PT can essentially change the behavior of the amplitude, as we have seen in our previous examples, and we hope that it takes place also in the four-fermion theory.

The summation of the leading logs is usually carried out within the renormalization group formalism that is absent in non-renormalizable models. However, in recent years we have developed such a formalism in the class of QFT models in higher dimensions, all of which are non-renormalizable. They include supersymmetric gauge theories~\cite{we1,we2}, scalar theories~\cite{scalar,eff}, and the four-dimensional supersymmetric Wess-Zumino model with a quartic superpotential~\cite{WZ}. It is tempting to  apply this formalism to the four-fermion interaction.

To do this, one first needs to calculate the leading diagrams for the scattering amplitudes in this theory to have a  solid playground and to understand the Lorentz structure of the amplitudes, since for fermions one can have different polarizations and different four-fermion operators.

In this paper we present our calculations of the on-shell $f f \rightarrow f f$ scattering amplitudes. These are precisely the amplitudes that we considered in our previous  papers. The complexity here is the spinor structure of the amplitudes.

We  start with the Lagrangian
\begin{equation}
{\cal L}=i \bar \Psi \hat \partial \Psi - \frac{G_F}{4} \bar \Psi \Psi \bar \Psi \Psi
\end{equation}
for the massless spinor field $\Psi$. This means that out of possible five operators: $ {\cal O}_i=1,\gamma^5,\gamma^\mu,\gamma^5\gamma^\mu,\gamma^\mu\gamma^\nu$, we take the simplest scalar one. We briefly consider the V-A  case at the end.
 In what follows we explore the two-component spinors (this is useful for massless fermions) and the spinor-helicity formalism and show that in the scalar case only two independent structures appear while in the V-A case one is left with a single structure.

We calculate the contributions to both amplitudes up to three loops but evaluate only the leading divergences. Our goal is to calculate the leading logarithmic asymptotic of the amplitude. In the dimensional regularization that we use everywhere the leading logs at $n$ loops, namely, $\log^n E^2/\mu^2$ are in a one-to-one correspondence with the  leading poles $\sim 1/\epsilon^n$. And to calculate the leading poles is much easier. So, in order to sum up the leading logs, one can actually sum up the leading poles and then make a replacement $1/\epsilon \to -\log E^2/\mu^2$ in the resulting expression.

Remarkably, the leading poles (leading logs) even in non-renormalizable models are subject to the renormalization group equations and can be calculated recurrently starting with the one loop without evaluation of the diagrams. This is achieved by establishing the recurrence relations which connect counterterms to the subsequent orders of perturbation theory.  We at the end present these relations and check that they indeed reproduce our perturbative calculations up to three loops. The derivation of recurrence relations is based on the general formalism of the R-operation and does not differ from what we did for other models; however, since in this particular case, we have two independent structures in all channels, they are quite cumbersome. For this reason we leave the derivation of recurrence relations and their analysis together with the corresponding RG equations to the next publication. Here we  concentrate on perturbative calculations of the diagrams.
 
 The paper is organized as follows: In Sec.2 we present the preliminaries of the Weyl spinors and the spinor-helicity formalism. In Sec.3 we describe the diagram calculation set up. Section 4 contains the calculation of the one-, two-, and three-loop scattering amplitudes. In Sec.5 we summarize our results for the scalar currents and in Sec.6 we consider the V-A interaction. Section 7 contains our conclusion.

\section{Preliminaries}

\subsection{Conventions and useful identities}

Throughout the paper we use the Minkowski metric
\begin{equation}
    g^{\mu\nu} = \text{diag}(1,-1,-1,-1),
\end{equation}
and the two-component Weyl spinors.
In the Weyl representation the $4 \times 4$ - Dirac matrices satisfying the anti-commutation relations
\begin{equation}
\{ \gamma^\mu, \gamma^\nu \} = 2g^{\mu\nu}I, \qquad
\{ \gamma^\mu, \gamma_5 \} = 0, \qquad
\gamma_5 = \frac{i}{24}\epsilon_{\mu\nu\rho\sigma}\gamma^\mu\gamma^\nu\gamma^\rho\gamma^\sigma,
\label{gamma}
\end{equation}
have the form
\begin{equation}
\gamma^{\mu} = \begin{pmatrix} 0 & \sigma^\mu \\ \bar{\sigma}^\mu & 0 \end{pmatrix}, \qquad  
\gamma_5 = i\gamma^0\gamma^1\gamma^2\gamma^3 = \begin{pmatrix} -I_2 & 0 \\ 0 & I_2 \end{pmatrix},
\label{Weyl}
\end{equation}
where the $2 \times 2$ sigma-matrices are defined by
\begin{equation}
    \sigma^\mu_{A\dot{B}} \equiv (1, \sigma^i), \qquad \bar{\sigma}^{\mu\dot{A}B} \equiv (1, -\sigma^i),
    \label{sigma}
\end{equation}
and $\sigma^i = (\sigma^1, \sigma^2, \sigma^3)$ are the standard Pauli matrices:
\begin{equation}
\sigma^1 = \begin{pmatrix} 0 & 1 \\ 1 & 0\end{pmatrix}, \qquad  
\sigma^2 = \begin{pmatrix} 0 & -i \\ i & 0\end{pmatrix}, \qquad  
\sigma^3 = \begin{pmatrix} 1 & 0 \\ 0 & -1\end{pmatrix}.
\end{equation}

The sigma-matrices satisfy the anticommutation and the Fierz identities
\begin{align}
   (&\sigma^{\mu}\bar{\sigma}^{\nu} + \sigma^{\nu}\bar{\sigma}^{\mu})_A^{\;\;B} = 2g^{\mu\nu}\delta_A^{\;\;B}, \\
   (&\bar{\sigma}^{\mu}\sigma^{\nu} + \bar{\sigma}^{\nu}\sigma^{\mu})_{\;\;\dot{A}}^{\dot{B}} = 2g^{\mu\nu}\delta_{\;\;\dot{A}}^{\dot{B}}, \\
    &\sigma^{\mu}_{A\dot{A}}\bar{\sigma}_{\mu}^{\dot{B}B} = 2\delta_A^{\;\;B}\delta_{\;\;\dot{A}}^{\dot{B}}, \label{FierzSigma} \\
    &\sigma^{\mu}_{A\dot{A}}\sigma_{\mu B\dot{B}} = 2\epsilon_{AB}\epsilon_{\dot{A}\dot{B}}, \\
    &\bar{\sigma}^{\mu\dot{A}A}\bar{\sigma}^{\mu\dot{B}B} = 2\epsilon^{\dot{A}\dot{B}}\epsilon^{AB},
\end{align}
where the 2-dimensional antisymmetric tensor is:
\begin{equation}
    \epsilon_{AB} = \begin{pmatrix} 0 & 1 \\ -1 & 0\end{pmatrix}, \qquad
    \epsilon_{BA} = - \epsilon_{AB}, \qquad
    \epsilon_{AB} = \epsilon_{\dot{A}\dot{B}} = \epsilon^{AB} = \epsilon^{\dot{A}\dot{B}},
    \label{epsilon}
\end{equation}
with the property
\begin{equation}
    \epsilon^{AC}\epsilon_{BC} = \delta^A_{\;\;B}, \qquad
    \epsilon^{\dot{A}\dot{C}}\epsilon_{\dot{B}\dot{C}} = \delta^{\dot{A}}_{\;\;\dot{B}}.
\end{equation}

This antisymmetric tensor is used  in raising and lowering the spinor index $A$ or $\dot{B}$, according to:
\begin{equation}
    \begin{aligned}
     &p^A = \epsilon^{AB}p_B, \qquad p_B = p^A \epsilon_{AB},  \\ 
     &p^{\dot{A}} = \epsilon^{\dot{A}\dot{B}}p_{\dot{B}}, \qquad 
      p_{\dot{B}} = p^{\dot{A}}\epsilon_{\dot{A}\dot{B}}.
    \end{aligned}
\end{equation}

\subsection{Dirac and Weyl spinors}

 Consider the general form of a free Dirac spinor field ~\cite{Elvang}
\begin{equation}
\Psi(x) = \sum_{s=\pm} \int \frac{d^3p}{(2\pi)^3 2E_p}\left[a_s(p)\,u_s(p)\,e^{ip.x} + a_s^\dagger(p)\,\upsilon_s(p)\,e^{-ip.x}\right],
\label{Psi}
\end{equation}
where $a_s(p)$ and $a_s^\dagger(p)$ are the fermionic creation and annihilation operators, $u_s(p)$ and $\upsilon_s(p)$ are 4-component spinors obeying the Dirac equation
\begin{equation}
(\not\!p+m)u(p)=0 \qquad \text{and} \qquad (\not\!p+m)\upsilon(p)=0.
\label{Deq}
\end{equation}
Here $\not\!p = p^{\mu} \gamma_{\mu}$ and we use the Dirac matrices $\gamma_{\mu}$ in the Weyl representation. The Dirac conjugated field is defined as usual 
\begin{equation}
\overline{\Psi}(x) = \Psi^\dagger \gamma^0, \qquad 
\gamma^0= \begin{pmatrix} 0 & I_2 \\ I_2 & 0 \end{pmatrix}.
\end{equation}
We associate $u(p)$ and $\overline{\upsilon}(p)$ with the wave functions of the incoming fermions and antifermions, respectively. 

These four-component Dirac spinors are constructed out of two Weyl spinors as follows\cite{1001, Schwinn, Dreiner}:
\begin{equation}
u(p) = \left(
\begin{array}{c}
u_R(p) \\ u_L(p)
\end{array}
\right) = \left(
\begin{array}{c}
p_A \\ p^{\dot B}
\end{array}
\right)  = \left(
\begin{array}{c}
|p\rangle \\ |p]
\end{array}
\right) 
\end{equation}
and
\begin{equation}
\overline{\upsilon}(p) = \left(
\begin{array}{cc}
\overline{\upsilon}_L(p) & \overline{\upsilon}_R(p)
\end{array}
\right) = \left(
\begin{array}{cc}
p^A & p_{\dot B}
\end{array}
\right)  = \left(
\begin{array}{cc}
\langle p| & [p|
\end{array}
\right)
\end{equation} 
Define now the chiral projection operators
\begin{equation}
P_R=\frac{1}{2}(1+\gamma_5)=\begin{pmatrix} 1 & 0 \\ 0 & 0 \end{pmatrix}, \qquad
P_L=\frac{1}{2}(1-\gamma_5)=\begin{pmatrix} 0 & 0 \\ 0 & 1 \end{pmatrix},
\end{equation}
with the following properties:
\begin{equation}
P_R^2=P_R, \qquad
P_L^2=P_L, \qquad
P_R P_L = P_L P_R = 0.
\label{proj}
\end{equation}

Then the external states of incoming particles  in terms of the two-component Weyl spinors are 
\begin{equation}
    \begin{aligned}
        P_R \, u(p) = u_R(p) = p_A = |p\rangle, \\
        P_L \, u(p) = u_L(p) = p^{\dot B} = |p], \\
        \overline{\upsilon}(p) \, P_R = \overline{\upsilon}_L(p) = p^A = \langle p|, \\
        \overline{\upsilon}(p) \, P_L = \overline{\upsilon}_R(p) = p_{\dot B} = [p|.
    \end{aligned}
    \label{sp}
\end{equation}

\subsection{Spinor - Helicity formalism} \label{shf}

Using the states (\ref{sp}), one can define the Lorentz - invariant spinor products with the following conventions:
\begin{equation}
\langle pk \rangle = p^A k_A = \epsilon^{AB} p_B k_A, \qquad
[pk] = p_{\dot{A}} k^{\dot{A}} = p^{\dot{B}} \epsilon_{\dot{B} \dot{A}} k^{\dot{A}},
\end{equation}
where the antisymmetric tensors $\epsilon^{AB}$ and $\epsilon_{\dot{B} \dot{A}}$ are defined  in (\ref{epsilon}). Then one has 
\begin{equation}
\langle pk \rangle [kp] = 2\, (p.k).
\label{Lorentz}
\end{equation}
The spinor products are antisymmetric 
\begin{equation}
\langle pk \rangle = - \langle kp \rangle, \qquad
[pk] = - [kp].
\end{equation}
All other products are equal to zero due to the properties of projection operators  (\ref{proj}):
\begin{equation}
\langle pk ] = \overline{\upsilon}(p)\, P_R P_L\, u(p) = 0, \qquad
[ pk \rangle = \overline{\upsilon}(p)\, P_L P_R\, u(p) = 0.
\end{equation}
For any alternating combination of $\sigma$ and $\overline{\sigma}$ matrices (it can be rather even or odd) one can prove that:
\begin{equation}
      \begin{aligned}
&\langle p |\underbrace{\sigma \overline{\sigma} ...}_\text{2n}| k \rangle = -\langle k |\sigma \overline{\sigma} ...| p \rangle, \hspace{1.8em}
\langle p |\underbrace{\sigma \overline{\sigma} ...}_\text{2n+1}| k \rangle = 0, \\
&[p |\underbrace{\overline{\sigma}\sigma  ...}_\text{2n}| k] = -[k |\overline{\sigma}\sigma  ...| p], \hspace{2.5em}
[p |\underbrace{\overline{\sigma}\sigma  ...}_\text{2n+1}| k] = 0, \\
&\langle p |\underbrace{\sigma \overline{\sigma} ...}_\text{2n}| k ] = [k |\underbrace{\overline{\sigma}\sigma  ...}_\text{2n}| p\rangle = 0, \quad
\langle p |\underbrace{\sigma \overline{\sigma} ...}_\text{2n+1}| k ] = [ k |\overline{\sigma} \sigma...| p \rangle.
\label{chains}
     \end{aligned}
\end{equation}
And finally, using (\ref{FierzSigma}), one can obtain the useful Fierz identity
\begin{equation}
\langle p_1|\sigma^\mu|p_2 ]\,[ p_3|\overline{\sigma}_\mu|p_4 \rangle = 2\, \langle p_1p_4 \rangle [p_3p_2]
\label{Fierz},
\end{equation}
which is used to reduce the Lorentz structure appearing in the diagrams of the basic set of amplitudes.

\section{The diagram calculation setup}

Consider now the scattering process $f f \rightarrow f f$. To calculate the Feynman diagrams corresponding to  this process, we follow the setup, which is described in \cite{Paraskevas}. We define the four-fermion interaction term in the Lagrangian as (one-flavor case)
\begin{equation}
    \mathcal{L}_{int} \equiv \Gamma^{(s_a,s_b)}\, \Gamma^{(s_c,s_d)} \bar{\psi}_{s_a}{\psi}_{s_b} \bar{\psi}_{s_c} {\psi}_{s_d},
\end{equation}
with $\Gamma$ being some arbitrary Dirac matrix structure. As was mentioned earlier, we start with the unit matrix $\Gamma^{(s_a,s_b)}=\Gamma^{(s_c,s_d)}=I$. Then considering all possible contractions with the interaction term in the first order of Dyson expansion, one obtains 
\begin{equation}
    \begin{aligned}
   \langle 0|(i \mathcal{L}_{int})b_4^{\dagger}d_3^{\dagger}b_2^{\dagger}d_1^{\dagger}|0 \rangle =  &\langle 0|(i \Gamma^{(s_a,s_b)}\, \Gamma^{(s_c,s_d)} \bar{\psi}_{s_a}{\psi}_{s_b} \bar{\psi}_{s_c} {\psi}_{s_d})b_4^{\dagger}d_3^{\dagger}b_2^{\dagger}d_1^{\dagger}|0 \rangle \\&
   = i\, \mathbf{\Gamma}^{(1234)} \bar{\upsilon}_1 u_2 \bar{\upsilon}_3 u_4, 
  \end{aligned}
   \label{int}
\end{equation}
where the vertex $\mathbf{\Gamma}^{(1234)}$ is
\begin{equation}
   i\, \mathbf{\Gamma}^{(1234)} \equiv 2i\left[\Gamma^{(1,2)}\, \Gamma^{(3,4)} - \Gamma^{(1,4)}\, \Gamma^{(3,2)}\right]. 
   \label{vertex}
\end{equation}

Diagrammatically it can be represented as
\begin{figure}[ht]
\center{\includegraphics[scale=0.3]{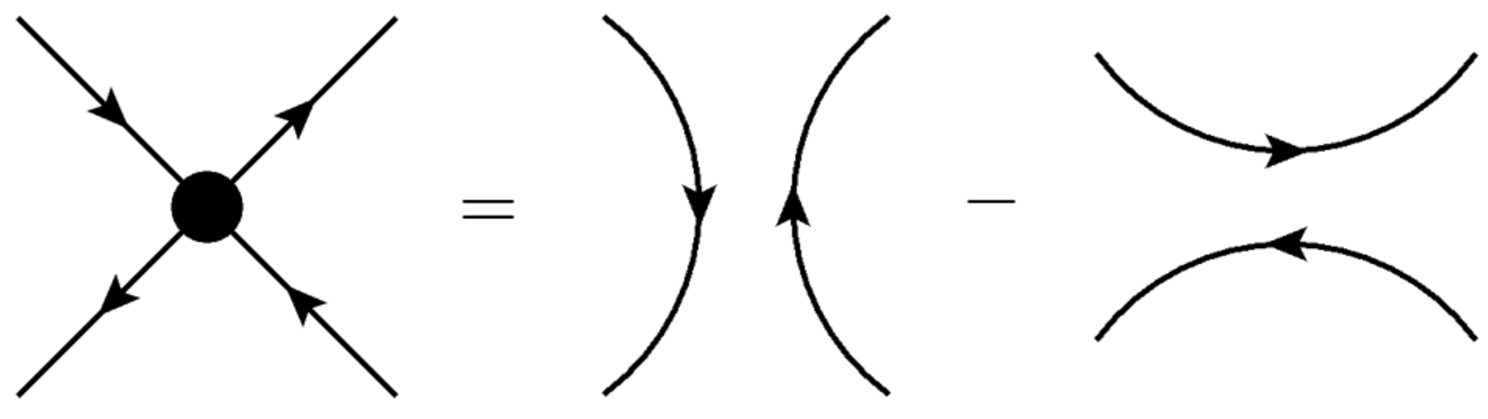}}
\end{figure}
 
Using the Weyl spinors and the spinor helicity formalism described above, these diagrams correspond to the following expressions:
\begin{equation}
    \includegraphics[scale = 0.15, trim=+15cm +6cm 0 0]{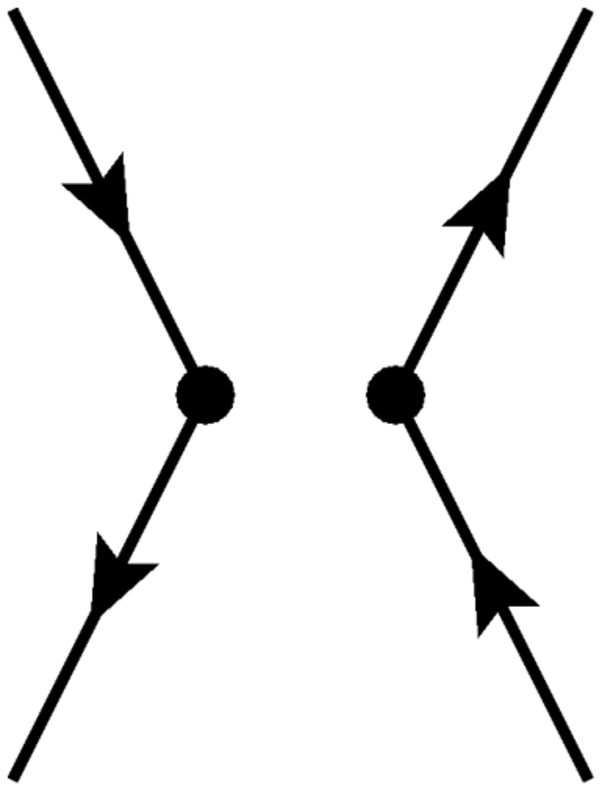} = \bar{\upsilon}_L(p_1) u_R(p_2) \bar{\upsilon}_R(p_3) u_L(p_4) = \langle 12 \rangle [34],
    \label{tree}
\end{equation}\\

\begin{equation}
    \includegraphics[scale = 0.15, trim=+9cm +4.5cm 0 0]{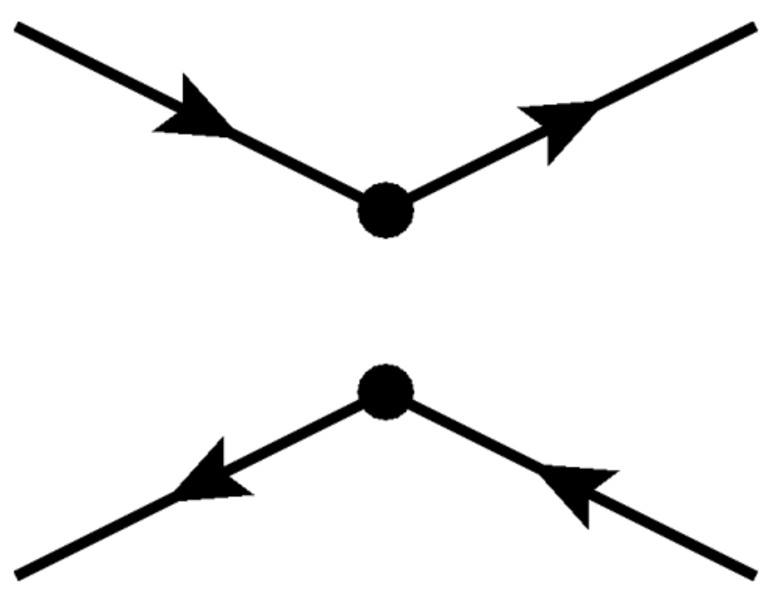} =- \bar{\upsilon}_L(p_1) u_R(p_4) \bar{\upsilon}_R(p_3) u_L(p_2) =- \langle 14 \rangle [32].
\end{equation}\\ \\
As will be shown below, only these two structures appear in all orders of the perturbation theory. 

In momentum space the propagator of the massless fermion reads
\begin{equation}
 \wick {\c\psi_2 \overline{\c\psi}_1} \rightarrow i \frac{k^\mu \cdot \sigma_\mu}{k^2} \equiv i D_{(2,1)}(p),
   \label{prop}
   \end{equation}
   
Here the propagator is defined for the case when the momenta and the fermion current flow in the same direction. The different direction case simply flips the sign of (\ref{prop}). 

\section{Loop corrections}

\subsection{One-loop}

\begin{figure}[ht]
\center{\includegraphics[scale=0.25]{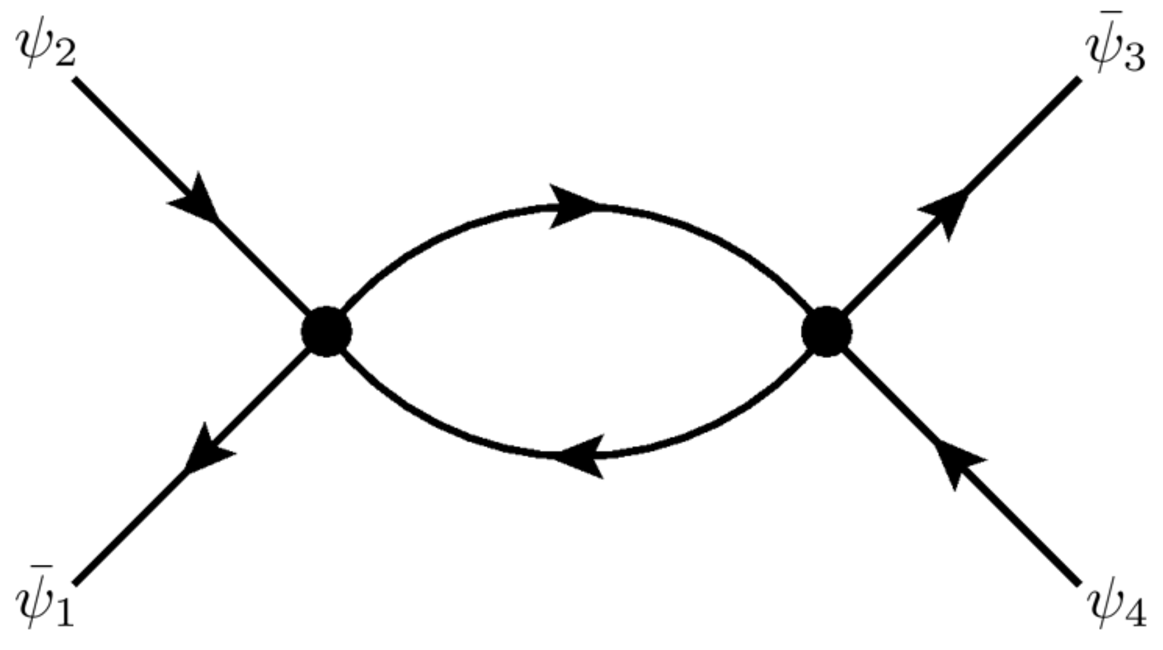}}
\caption{The s-channel bubble diagram of the first order}
\label{Bubs1}
\end{figure}

The first diagram is the s-channel bubble shown in Fig.\ref{Bubs1},
which can be represented schematically by the multiplication of two trees:
\begin{figure}[ht]
\center{\includegraphics[scale=0.25]{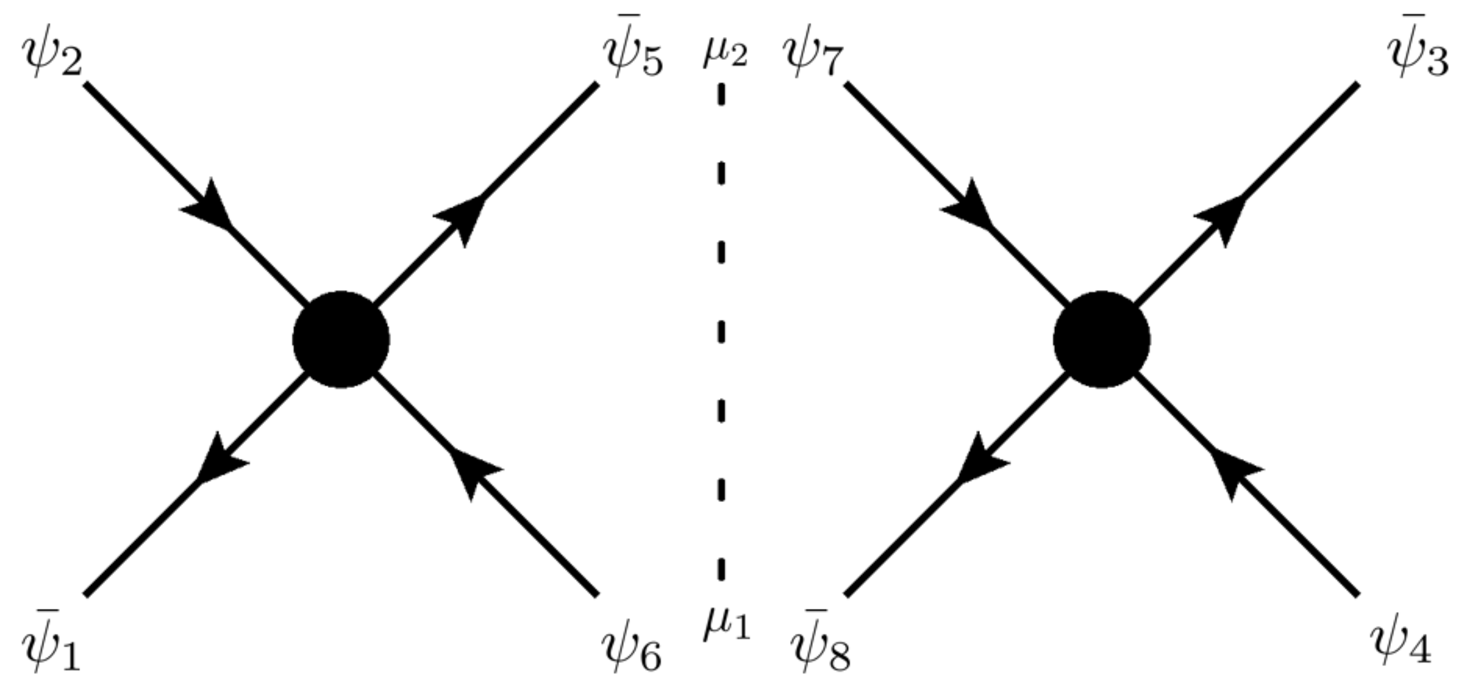}}
\end{figure}\\
Using (\ref{int}) and (\ref{vertex}), one can write down the expression for the s-channel diagram as
\begin{equation}
      \begin{aligned}  
   sBub_1&=A_4^{(0)}(\bar{\psi}_1 \psi_2 \bar{\psi}_5 \psi_6) \times A_4^{(0)}(\bar{\psi}_8 \psi_7 \bar{\psi}_3 \psi_4) =\\&= -i^2\left[\Gamma^{(1,2)}\Gamma^{(5,6)}-\Gamma^{(1,6)}\Gamma^{(5,2)}\right]\left[\Gamma^{(8,7)}\Gamma^{(3,4)}-\Gamma^{(8,4)}\Gamma^{(3,7)}\right]\times \\&\times D^{\mu_1}_{(6,8)}(k) D^{\mu_2}_{(7,5)}(k+p)\, \bar{\upsilon}_L(p_1) u_R(p_2) \bar{\upsilon}_R(p_3) u_L(p_4).
      \end{aligned}
\end{equation}

This leads to the following four contribution terms:
\begin{eqnarray*}
        1) &-\left(\bar{\upsilon}_L(p_1) \Gamma^{(1,2)} u_R(p_2)\right) \left(\Gamma^{(5,6)} D^{\mu_1}_{(6,8)} \Gamma^{(8,7)} D^{\mu_2}_{(7,5)}\right) \left(\bar{\upsilon}_R(p_3) \Gamma^{(3,4)} u_L(p_4)\right) =\\ &=  -\langle 12 \rangle\, Tr [\sigma^{\mu_1} \bar{\sigma}^{\mu_2}] [34] I_1^{\mu_1\mu_2} ,\\
        2) &\left(\bar{\upsilon}_L(p_1) \Gamma^{(1,6)} D^{\mu_1}_{(6,8)} \Gamma^{(8,7)} D^{\mu_2}_{(7,5)} \Gamma^{(5,2)} u_R(p_2)\right) \left(\bar{\upsilon}_R(p_3) \Gamma^{(3,4)} u_L(p_4)\right)=\\ &= \langle 1|\sigma^{\mu_1} \bar{\sigma}^{\mu_2}|2\rangle [34] I_1^{\mu_1\mu_2},\\
        3) &\left(\bar{\upsilon}_L(p_1) \Gamma^{(1,2)} u_R(p_2)\right) \left(\bar{\upsilon}_R(p_3) \Gamma^{(3,7)} D^{\mu_2}_{(7,5)} \Gamma^{(5,6)} D^{\mu_1}_{(6,8)} \Gamma^{(8,4)} u_L(p_4)\right)=\\ &= \langle 12\rangle [3|\sigma^{\mu_2} \bar{\sigma}^{\mu_1}|4] I_1^{\mu_1\mu_2},\\
        4) &-\left(\bar{\upsilon}_L(p_1) \Gamma^{(1,6)} D^{\mu_1}_{(6,8)} \Gamma^{(8,4)} u_L(p_4)\right) \left(\bar{\upsilon}_R(p_3) \Gamma^{(3,7)} D^{\mu_2}_{(7,5)} \Gamma^{(5,2)} u_R(p_2)\right)=\\ &=-\langle 1|\sigma^{\mu_1}|4][3|\bar{\sigma}^{\mu_2}|2\rangle I_1^{\mu_1\mu_2}.
\end{eqnarray*}
Here $I_1^{\mu\nu}$ is the one-loop divergent integral
\begin{equation}
    I_1^{\mu\nu} = \int d^4k \frac{k^{\mu}(k+p)^{\nu}}{k^2 (k+p)^2} .
\end{equation}
Taking the divergent part calculated within the dimensional regularization one gets
\begin{equation}
Div \ I_1^{\mu\nu} = - \frac{1}{6\epsilon}\left(p^{\mu} p^{\nu} + \frac{(p.p)}{2}g^{\mu \nu}\right),
\end{equation}
which gives the following four contributions:
 \begin{equation*}
    \begin{aligned} 
 1)&\quad\includegraphics[scale = 0.15, trim=0 +2cm 0 0]{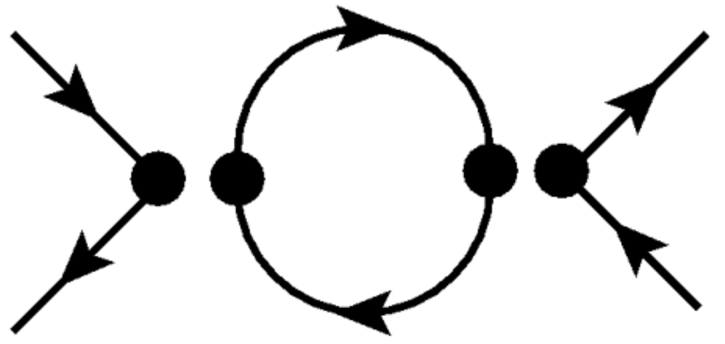} ~ = \frac{s}{\epsilon}  \langle 12 \rangle [34],\\ \\
 2)&\quad\includegraphics[scale = 0.15, trim=0 +2.5cm 0cm 0]{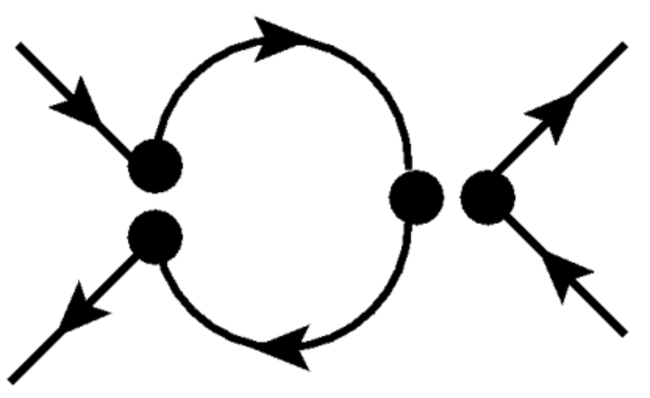} ~~ =  -\frac{s}{2\epsilon} \langle 12 \rangle [34],\\ \\
 3)&\quad\includegraphics[scale = 0.15, trim=0 +2.5cm 0cm 0]{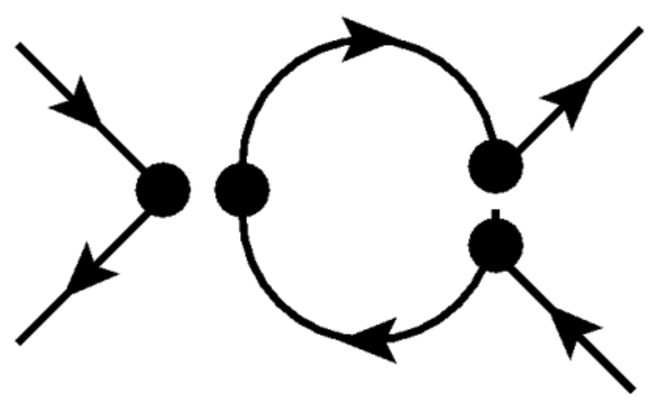} ~~ = -\frac{s}{2\epsilon}  \langle 12 \rangle [34],\\ \\
 4)&\quad\includegraphics[scale = 0.15, trim=0 +2.5cm 0cm 0]{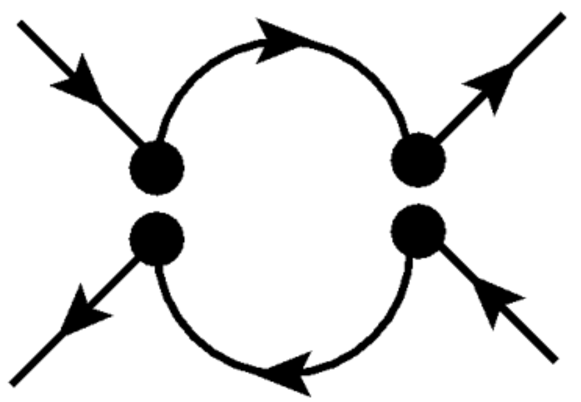} ~~~ = \frac{1}{6\epsilon}\left(\langle 1| \slashed{p}|4] [3|\slashed{p}|2 \rangle + \frac{s}{2}\langle 1|\sigma^{\mu_1}|4] [3|\bar{\sigma}_{\mu_1}|2 \rangle\right).
 \end{aligned}
\end{equation*}\\
 
In the fourth expression we use the momentum conservation $p=p_1+p_2=-p_3-p_4$ for the first term and the Fierz identity (\ref{Fierz}) for the second term resulting in
\begin{equation*}
    \begin{aligned}
&\frac{1}{6\epsilon}\left(-\langle 1|(p_1+p_2)|4] [3|(p_3+p_4)|2 \rangle + \frac{s}{2}\langle 1|\sigma^{\mu_1}|4] [3|\bar{\sigma}_{\mu_1}|2 \rangle\right) =\\ &= \frac{1}{6\epsilon}\left(-\langle 1|2|4] [3|4|2 \rangle + s\langle 12 \rangle [34] \right) = \frac{1}{6\epsilon}\left(-\langle 12 \rangle [24] [34] \langle 42 \rangle + s\langle 12 \rangle [34] \right).
    \end{aligned}
\end{equation*}

Finally, using (\ref{Lorentz}), we come to 
\begin{equation*}
4) \frac{1}{6\epsilon}\left(-\langle 12 \rangle [24] [34] \langle 42 \rangle + s\langle 12 \rangle [34] \right) = \frac{1}{6\epsilon}\left(-u\langle 12 \rangle [34] + s\langle 12 \rangle [34] \right) = \frac{s-u}{6\epsilon}\langle 12 \rangle [34].
\end{equation*}
Summing up all four answers one gets 
\begin{equation}
sBub_1=\frac{s-u}{6\epsilon}\langle 12 \rangle [34].
\end{equation}

Note that the usage of the Weyl spinors gives an essential simplification compared to the Dirac ones.
 
\begin{figure}[ht]
\center{\includegraphics[scale=0.25]{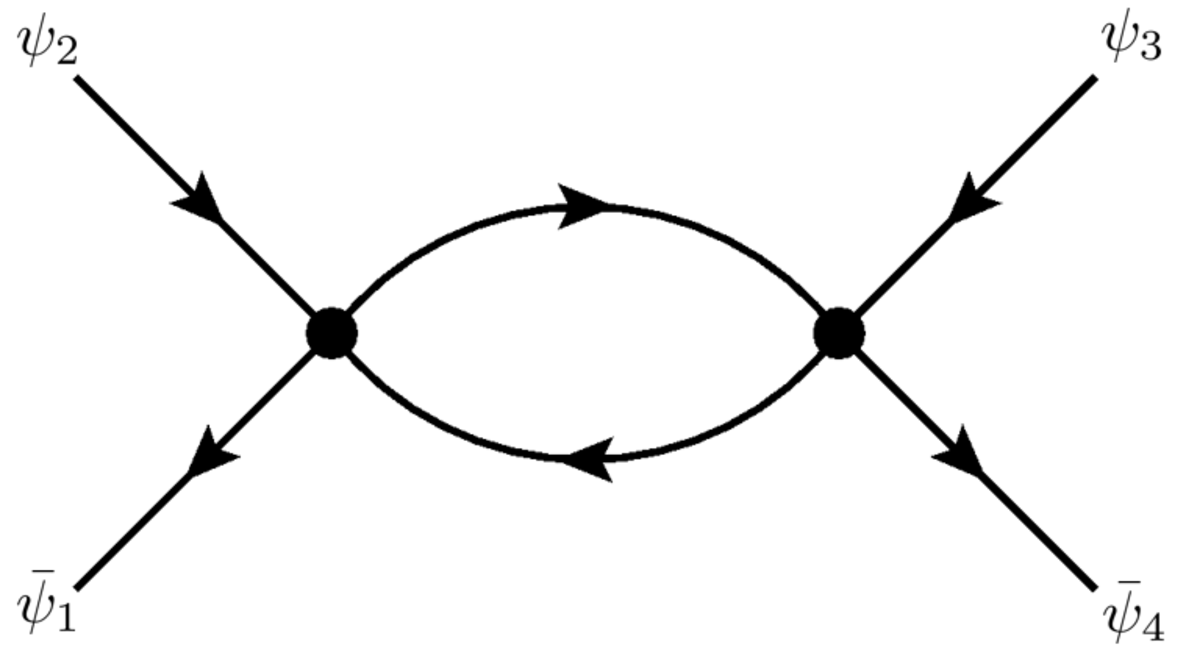}}
\caption{The second diagram in the s-channel}
\label{Bubs2}
\end{figure}

The other s-channel diagram  (Fig.\ref{Bubs2}) can be obtained if we set particle 3 to be  a fermion and particle 4 to be  an anti-fermion. The only contributing term is
 \begin{equation*}
 \quad\includegraphics[scale = 0.15, trim=0 +2.5cm 0cm 0]{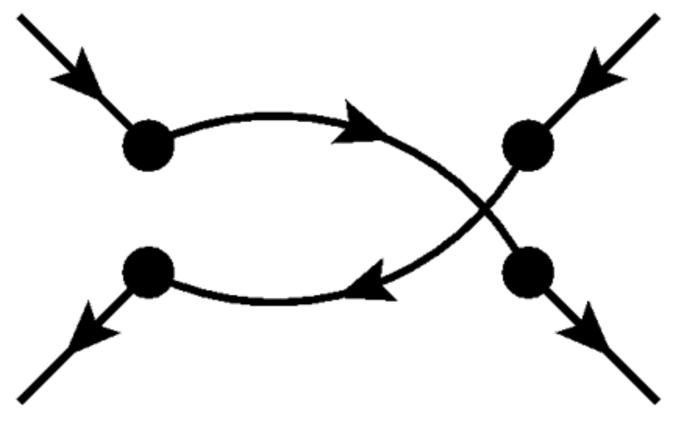} ~~ = -\frac{1}{6\epsilon}\left(\langle 1|\slashed{p}|3] [4|\slashed{p}|2 \rangle + \frac{s}{2}\langle 1|\sigma^{\mu_1}|3] [4|\bar{\sigma}_{\mu_1}|2 \rangle\right).
 \end{equation*}\\

Applying again the momentum conservation and the Fierz identity, we obtain
\begin{equation}
sBub_2=\frac{s-t}{6\epsilon}\langle 12 \rangle [34].
\end{equation}

The full contribution to the s - channel is then
\begin{equation}
RS_1 = sBub_1 + sBub_2 =\left(\frac{s-u}{6\epsilon} + \frac{s-t}{6\epsilon}\right)\langle 12 \rangle [34] = \frac{s}{2\epsilon} \langle 12 \rangle[34].
\end{equation}

\begin{figure}[ht]
\center{\includegraphics[scale=0.25]{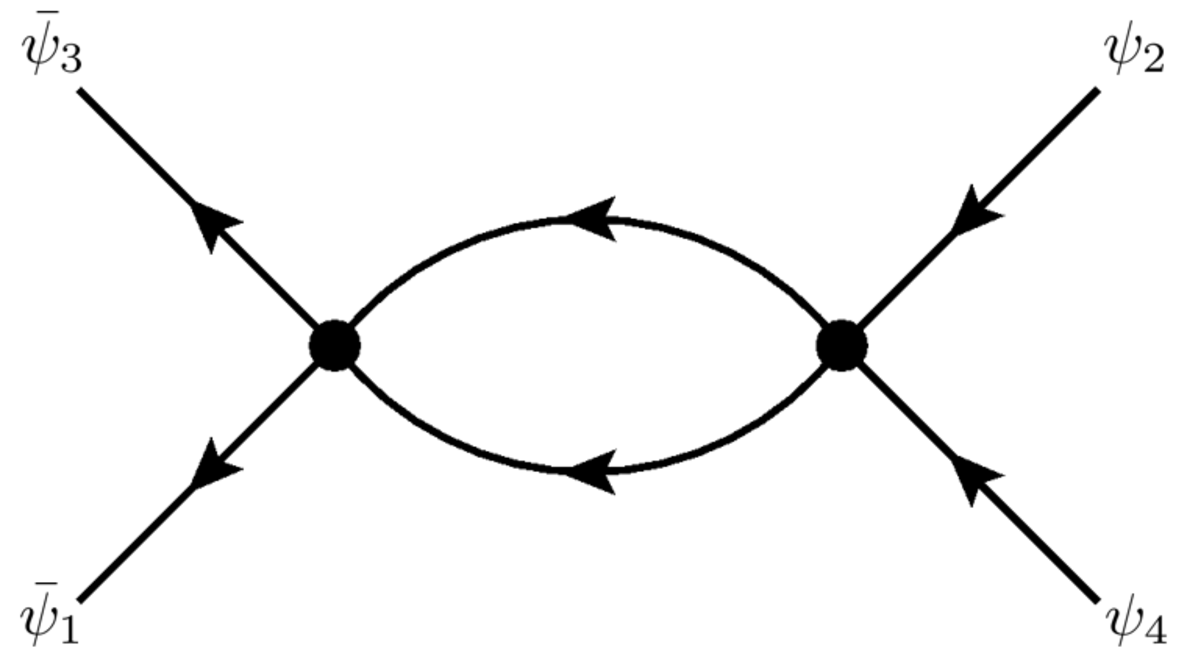}}
\caption{The u-channel diagram in the first order}
\label{Bubu1}
\end{figure}

The next diagram to be considered is the u-channel diagram (Fig.\ref{Bubu1}).

The multiplication of the tree level diagrams is not trivial here, since in the u-channel diagram the external legs interlace. It can be represented schematically as
\begin{figure}[ht]
\center{\includegraphics[scale=0.25]{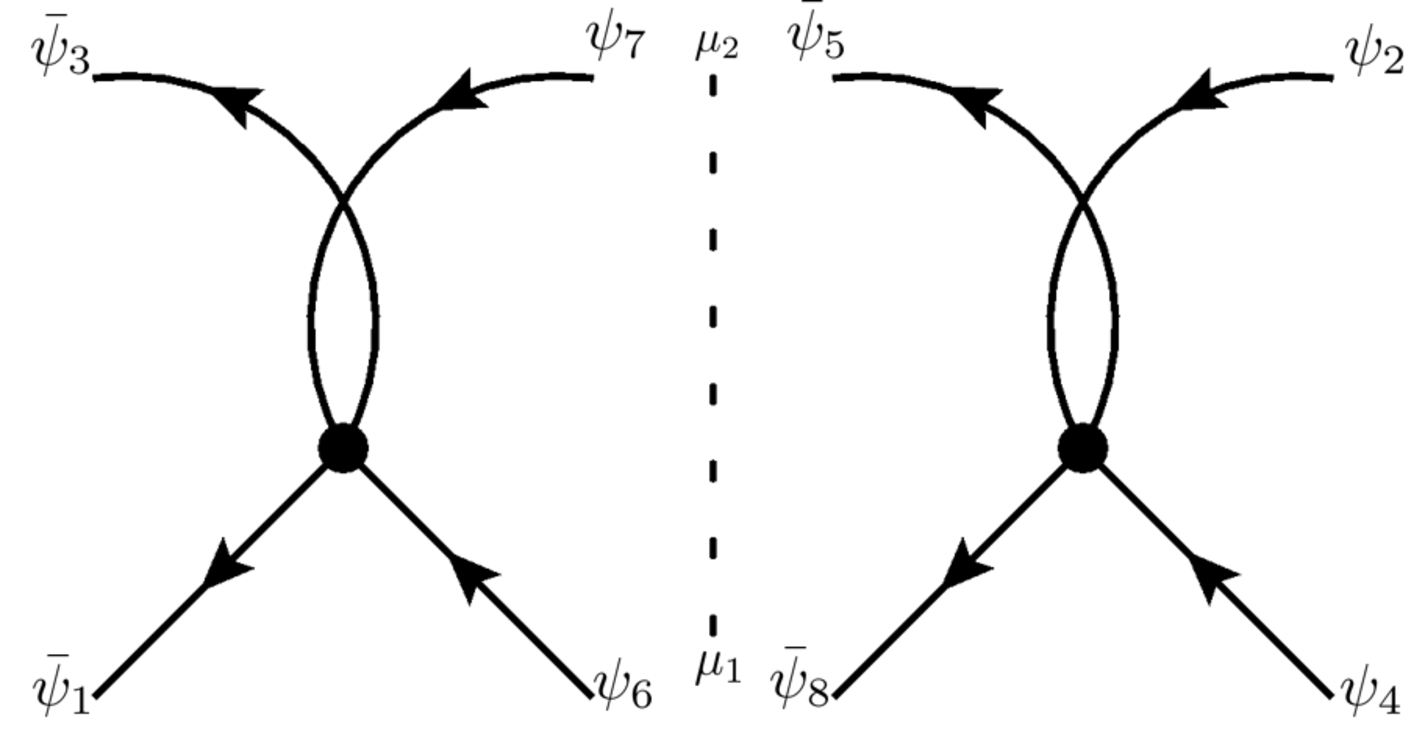}}
\end{figure}

The corresponding expression reads
\begin{equation}
      \begin{aligned}  
  uBub =  &A_4^{(0)}(\bar{\psi}_1 \psi_7 \bar{\psi}_3 \psi_6) \times A_4^{(0)}(\bar{\psi}_8 \psi_2 \bar{\psi}_5 \psi_4) =  i^2\left[\Gamma^{(1,7)}\Gamma^{(3,6)}-\Gamma^{(1,6)}\Gamma^{(3,7)}\right]\times \\ &\times\left[\Gamma^{(8,2)}\Gamma^{(5,4)}-\Gamma^{(8,4)}\Gamma^{(5,2)}\right] D^{\mu_1}_{(6,8)}(k) D^{\mu_2}_{(7,5)}(k+p)\, \bar{\upsilon}_L(p_1) u_R(p_2) \bar{\upsilon}_R(p_3) u_L(p_4).
      \end{aligned}
\end{equation}
Again, we come to four terms:
\begin{equation*}
    \begin{aligned}
        1)\, &-\left(\bar{\upsilon}_L(p_1) \Gamma^{(1,7)} D^{\mu_2}_{(7,5)} \Gamma^{(5,2)} u_R(p_2)\right) \left(\bar{\upsilon}_R(p_3) \Gamma^{(3,6)} D^{\mu_1}_{(6,8)} \Gamma^{(8,4)} u_L(p_4)\right)=\\ &=-\langle 1|\sigma^{\mu_2}|2\rangle [3|\bar{\sigma}^{\mu_1}|4] I_1^{\mu_1\mu_2} \Rightarrow 0,\\
        2)\, &\left(\bar{\upsilon}_L(p_1) \Gamma^{(1,7)} D^{\mu_2}_{(7,5)} \Gamma^{(5,4)} u_L(p_4)\right) \left(\bar{\upsilon}_R(p_3) \Gamma^{(3,6)} D^{\mu_1}_{(6,8)} \Gamma^{(8,2)} u_R(p_2)\right)=\\ &=\langle 1|\sigma^{\mu_2}|4][3|\bar{\sigma}^{\mu_1}|2\rangle I_1^{\mu_1\mu_2} \Rightarrow -\frac{u}{3\epsilon} \langle 12 \rangle [34],\\
        3)\, &\left(\bar{\upsilon}_L(p_1) \Gamma^{(1,6)} D^{\mu_1}_{(6,8)} \Gamma^{(8,4)} u_L(p_4)\right) \left(\bar{\upsilon}_R(p_3) \Gamma^{(3,7)} D^{\mu_2}_{(7,5)} \Gamma^{(5,2)} u_R(p_2)\right)=\\ &=\langle 1|\sigma^{\mu_1}|4][3|\bar{\sigma}^{\mu_2}|2\rangle I_1^{\mu_1\mu_2} \Rightarrow -\frac{u}{3\epsilon}\langle 12 \rangle [34],\\
        4)\, &-\left(\bar{\upsilon}_L(p_1) \Gamma^{(1,6)} D^{\mu_1}_{(6,8)} \Gamma^{(8,2)} u_R(p_2)\right) \left(\bar{\upsilon}_R(p_3) \Gamma^{(3,7)} D^{\mu_2}_{(7,5)} \Gamma^{(5,4)} u_L(p_4)\right)=\\ &=-\langle 1|\sigma^{\mu_1}|2\rangle [3|\bar{\sigma}^{\mu_2}|4] I_1^{\mu_1\mu_2} \Rightarrow 0,
    \end{aligned}
\end{equation*}
where $1)$ and $4)$ are equal to zero due to (\ref{chains}).

Summing up, one gets in the u-channel
\begin{equation}
RU_1=uBub=\frac{1}{2}\left(-\frac{u}{3\epsilon}-\frac{u}{3\epsilon}\right)\langle 12 \rangle [34] = -\frac{u}{3\epsilon} \langle 12 \rangle [34].
\end{equation}

\begin{figure}[ht]
\center{\includegraphics[scale=0.25]{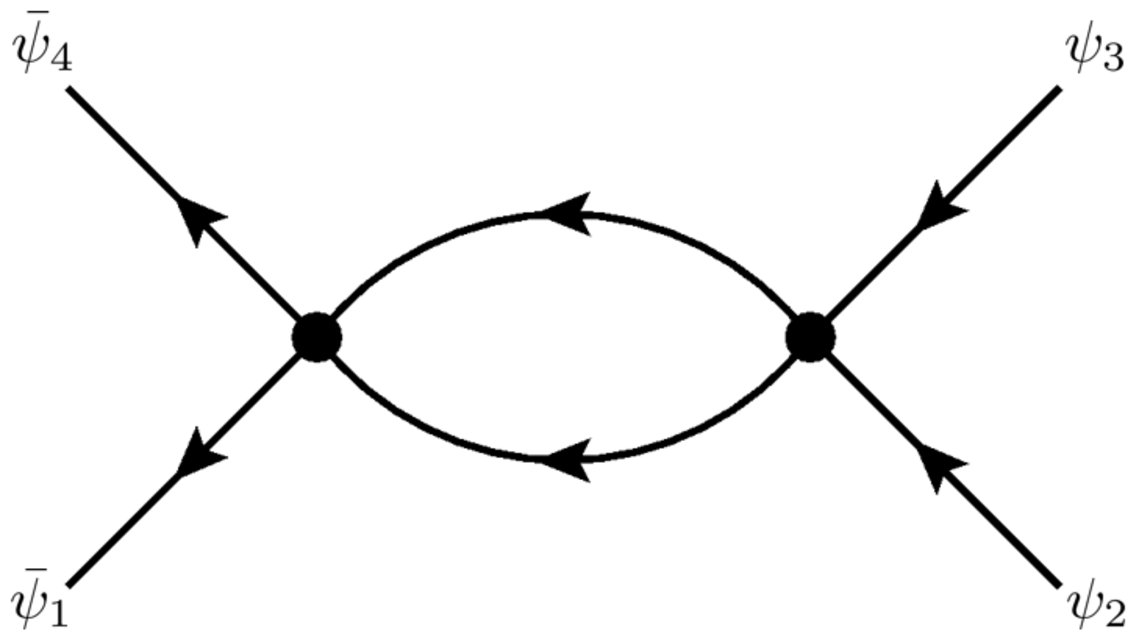}}
\caption{The t-channel diagram}
\label{Bubt1}
\end{figure}

The last one is the t-channel diagram shown in Fig.\ref{Bubt1}
\noindent which yields
\begin{equation}
RT_1=tBub = -\frac{t}{3\epsilon} \langle 12 \rangle [34].
\end{equation}

Thus, the total one-loop contribution to the $\langle 12 \rangle [34]$ part of the amplitude is (using $u = -s -t$):
\begin{equation}
 R_1(s,t) = RS_1+RT_1+RU_1 =\frac{5s}{6\epsilon} \langle 12 \rangle [34].
 \label{R1}
\end{equation}

\begin{figure}[ht]
\center{\includegraphics[scale=0.5, trim=+2cm 0cm 0cm 0]{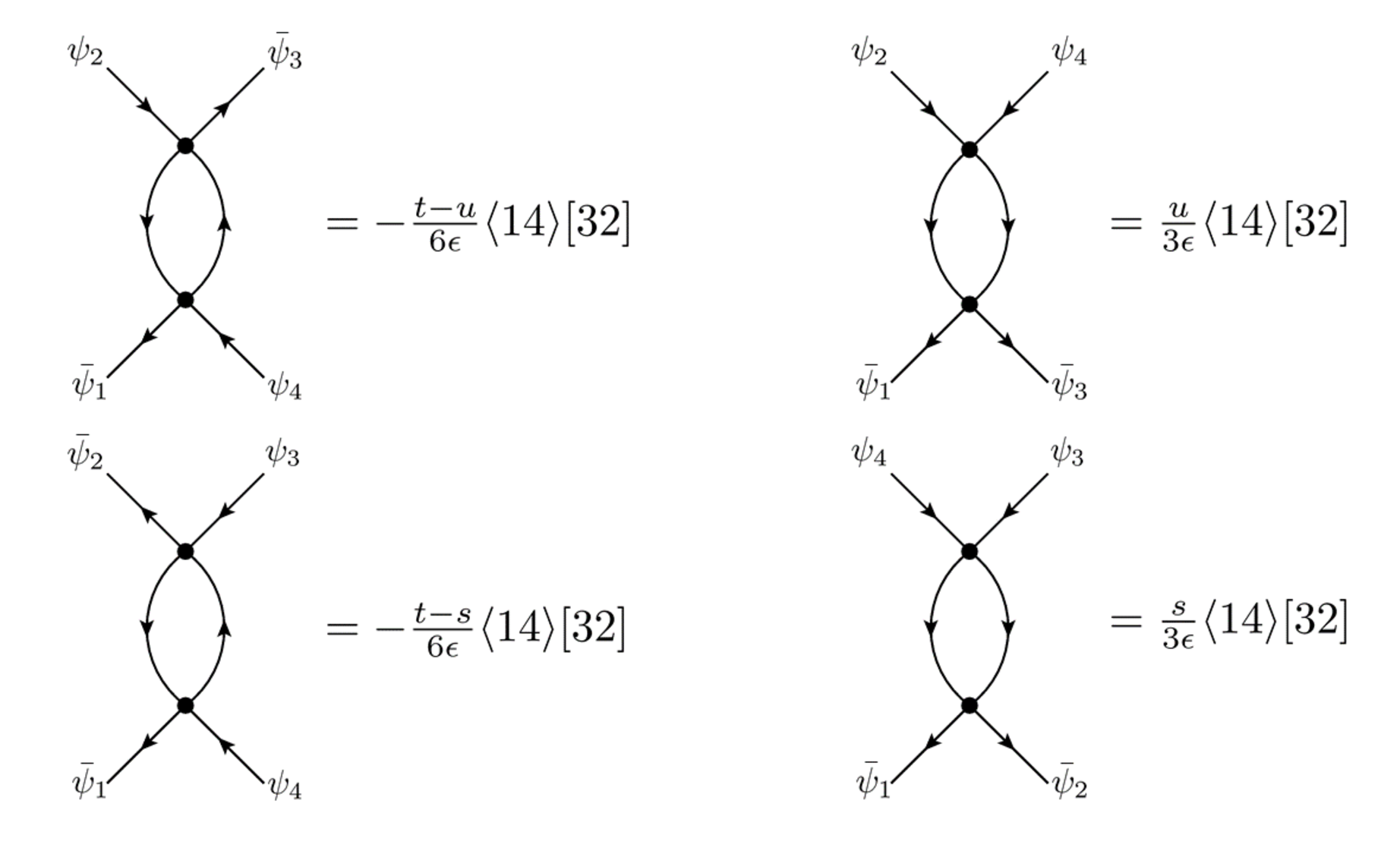}}
\caption{Diagrams and their contributions to $\langle 14 \rangle [32]$}
\label{BubstuL}
\end{figure}

The other part of the amplitude, which is proportional to $\langle 14 \rangle [32]$, includes the diagrams
shown in Fig.\ref{BubstuL}.
\\
They are summed up to
\begin{equation}
 L_1(t,s) =-\frac{5t}{6\epsilon} \langle 14 \rangle [32].
\end{equation}\\

For the total one-loop amplitude we obtain 
\begin{equation}
A_4^{(1)} = R_1(s,t) + L_1(t,s) = \frac{5s}{6\epsilon} \langle 12 \rangle [34] - \frac{5t}{6\epsilon} \langle 14 \rangle [32].
\end{equation}
This has to be compared with expression for  the tree-level amplitude
\begin{equation}
A_4^{(0)} = \langle 12 \rangle [34] - \langle 14 \rangle [32].
\end{equation}

Therefore one  can obtain the answer for  the $\langle 14 \rangle [32]$ part from the $\langle 12 \rangle [34]$ one by interchanging $p_2 \leftrightarrow p_4$ and keeping in mind the anticommutation sign. This rule applies to all orders of the perturbation theory.

\subsection{Two loops}

In two loops there are two topologically distinct diagrams contributing to the  $2\times 2$ amplitude. Consider first the double bubble one in the s - channel
\begin{figure}[ht]
\center{\includegraphics[scale=0.25]{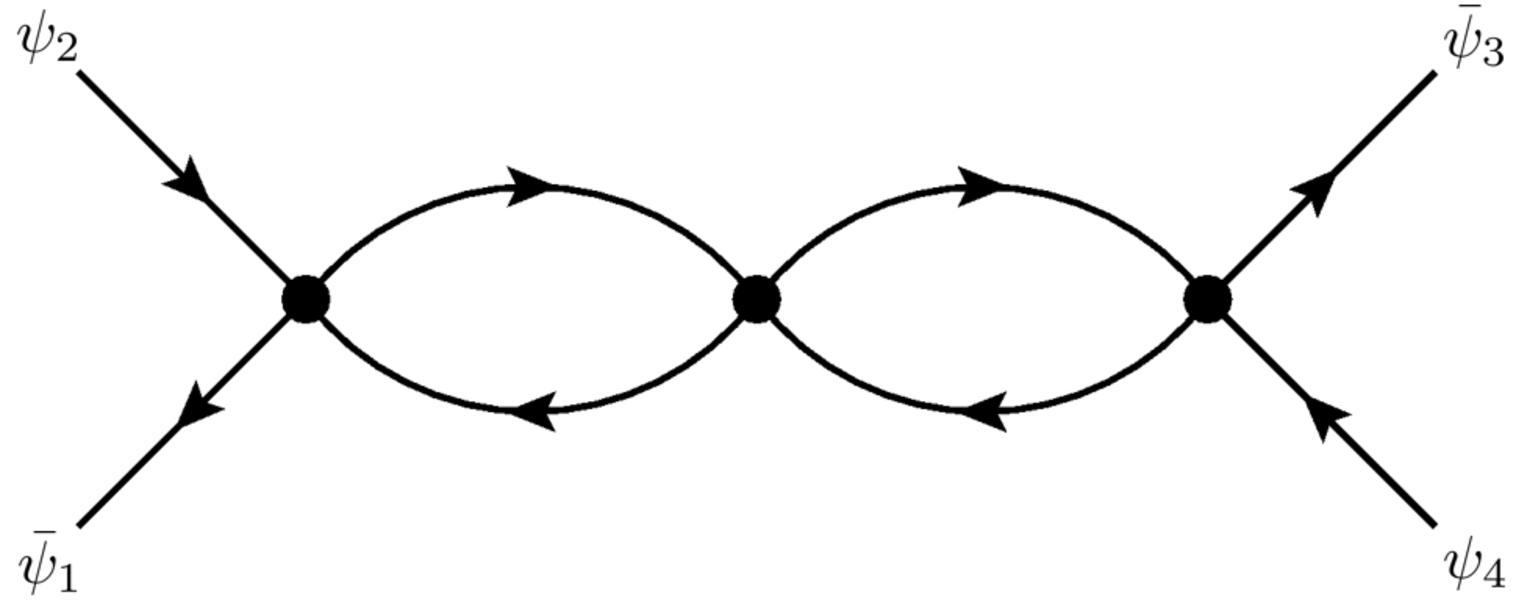}}
\caption{The s-channel double bubble topology}
\label{DBubs}
\end{figure}

The corresponding expression  can be obtained multiplying three tree level
diagrams
\begin{equation}
      \begin{aligned}  
 sDBub = &A_4^{(0)}(\bar{\psi}_1 \psi_2 \bar{\psi}_5 \psi_6) \times A_4^{(0)}(\bar{\psi}_8 \psi_7 \bar{\psi}_9 \psi_{10}) \times A_4^{(0)}(\bar{\psi}_{12} \psi_{11} \bar{\psi}_3 \psi_4) \\&\hspace*{-2cm}\ =i^3\left[\Gamma^{(1,2)}\Gamma^{(5,6)}-\Gamma^{(1,6)}\Gamma^{(5,2)}\right]\left[\Gamma^{(8,7)}\Gamma^{(9,10)}-\Gamma^{(8,10)}\Gamma^{(9,7)}\right]\left[\Gamma^{(12,11)}\Gamma^{(3,4)}-\Gamma^{(12,4)}\Gamma^{(3,11)}\right] \\ &\hspace*{-1cm}\times D^{\mu_1}_{(6,8)}(k) D^{\mu_2}_{(7,5)}(k+p) D^{\mu_3}_{(10,12)}(l) D^{\mu_4}_{(11,9)}(l+p)\, \bar{\upsilon}_L(p_1) u_R(p_2) \bar{\upsilon}_R(p_3) u_L(p_4).
     \end{aligned}
\end{equation}

In this case, one has eight terms:
\begin{equation*}
    \begin{aligned}
        &1)\, \langle 12 \rangle Tr[\sigma^{\mu_1} \bar{\sigma}^{\mu_2}] Tr[\sigma^{\mu_3} \bar{\sigma}^{\mu_4}] [34] I_{2,1}^{\mu_1\mu_2\mu_3\mu_4} \Rightarrow \frac{s^2}{\epsilon^2}, \\
        &2)\, -\langle 1|\sigma^{\mu_1} \bar{\sigma}^{\mu_2}|2 \rangle Tr[\sigma^{\mu_3} \bar{\sigma}^{\mu_4}] [34] I_{2,1}^{\mu_1\mu_2\mu_3\mu_4} \Rightarrow -\frac{s^2}{2\epsilon^2}, \\
        &3)\, -\langle 12 \rangle Tr[\sigma^{\mu_1} \bar{\sigma}^{\mu_3} \sigma^{\mu_4} \bar{\sigma}^{\mu_2}] [34] I_{2,1}^{\mu_1\mu_2\mu_3\mu_4} \Rightarrow -\frac{s^2}{2\epsilon^2}, \\
        &4)\, \langle 1|\sigma^{\mu_1} \bar{\sigma}^{\mu_3} \sigma^{\mu_4} \bar{\sigma}^{\mu_2}|2 \rangle [34] I_{2,1}^{\mu_1\mu_2\mu_3\mu_4}\Rightarrow \frac{s^2}{4\epsilon^2}, \\
        &5)\, -\langle 12 \rangle Tr[\sigma^{\mu_1} \bar{\sigma}^{\mu_2}] [3|\bar{\sigma}^{\mu_4} \sigma^{\mu_3}|4] I_{2,1}^{\mu_1\mu_2\mu_3\mu_4} \Rightarrow -\frac{s^2}{2\epsilon^2},\\
        &6)\, \langle 1|\sigma^{\mu_1} \bar{\sigma}^{\mu_2}|2 \rangle [3|\bar{\sigma}^{\mu_4} \sigma^{\mu_3}|4] I_{2,1}^{\mu_1\mu_2\mu_3\mu_4} \Rightarrow\frac{s^2}{4\epsilon^2}, \\
        &7)\, \langle 12 \rangle [3|\bar{\sigma}^{\mu_4} \sigma^{\mu_2} \bar{\sigma}^{\mu_1} \sigma^{\mu_3}|4] I_{2,1}^{\mu_1\mu_2\mu_3\mu_4} \Rightarrow \frac{s^2}{4\epsilon^2},\\
        &8)\, -\langle 1|\sigma^{\mu_1} \bar{\sigma}^{\mu_3}|4 ] [3|\bar{\sigma}^{\mu_4} \sigma^{\mu_2}|2\rangle I_{2,1}^{\mu_1\mu_2\mu_3\mu_4} \Rightarrow 0,\\
    \end{aligned}
\end{equation*}
where $I_{2,1}^{\mu_1\mu_2\mu_3\mu_4}$ is the two-loop bubble integral
\begin{equation}
I_{2,1}^{\mu_1\mu_2\mu_3\mu_4}=I_1^{\mu_1\mu_2}I_1^{\mu_3\mu_4}.
\end{equation}
% We evaluate this integral by using the R' - operation\cite{BSh, Kazakov}.
% \begin{figure}[ht]
% \center{\includegraphics[scale=0.45]{images/Figure_7.eps}}
% \caption{Subtraction of divergent subgraphs following the R' - operation}
% \label{R}
% \end{figure}\\

Summing up, one gets
\begin{equation}
   sDBub=\frac{s^2}{4\epsilon^2} \langle 12 \rangle [34].
\end{equation}\\

\begin{figure}[h]
\center{\includegraphics[scale=0.25]{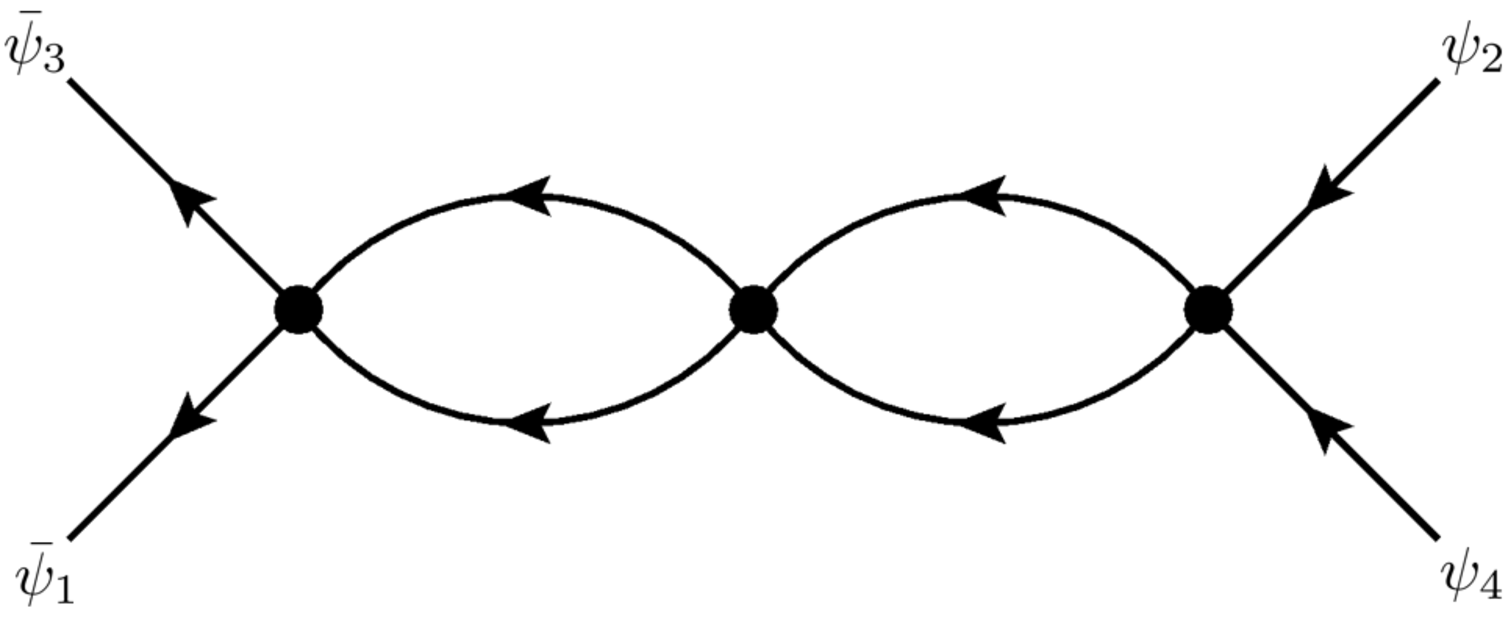}}
\caption{Double bubble in the u-channel}
\label{DBubu}
\end{figure}

Consider now the u-channel diagram shown in Fig.\ref{DBubu}.

One has the following contributions:
\begin{equation*}
    \begin{aligned}
        &1)\, \langle 1|\sigma^{\mu_2} \bar{\sigma}^{\mu_4}|2 \rangle [3|\bar{\sigma}^{\mu_1} \sigma^{\mu_3}|4] I_{2,1}^{\mu_1\mu_2\mu_3\mu_4} \Rightarrow 
        \frac{u^2}{9\epsilon^2}\langle 12 \rangle [34],\\
        &2)\, -\langle 1|\sigma^{\mu_2} \bar{\sigma}^{\mu_3}|4] [3|\bar{\sigma}^{\mu_1} \sigma^{\mu_4}|2\rangle I_{2,1}^{\mu_1\mu_2\mu_3\mu_4} \Rightarrow 
        0,\\
        &3)\, -\langle 1|\sigma^{\mu_2} \bar{\sigma}^{\mu_4}|4] [3|\bar{\sigma}^{\mu_1} \sigma^{\mu_3}|2\rangle I_{2,1}^{\mu_1\mu_2\mu_3\mu_4} \Rightarrow 
        0,\\
        &4)\, \langle 1|\sigma^{\mu_2} \bar{\sigma}^{\mu_3}|2 \rangle [3|\bar{\sigma}^{\mu_1} \sigma^{\mu_4}|4] I_{2,1}^{\mu_1\mu_2\mu_3\mu_4} \Rightarrow 
        \frac{u^2}{9\epsilon^2}\langle 12 \rangle [34],\\
        &5)\, -\langle 1|\sigma^{\mu_1} \bar{\sigma}^{\mu_3}|4] [3|\bar{\sigma}^{\mu_2} \sigma^{\mu_4}|2\rangle I_{2,1}^{\mu_1\mu_2\mu_3\mu_4} \Rightarrow 
        0,\\
        &6)\, \langle 1|\sigma^{\mu_1} \bar{\sigma}^{\mu_4}|2 \rangle [3|\bar{\sigma}^{\mu_2} \sigma^{\mu_3}|4] I_{2,1}^{\mu_1\mu_2\mu_3\mu_4} \Rightarrow 
        \frac{u^2}{9\epsilon^2}\langle 12 \rangle [34],\\
        &7)\, \langle 1|\sigma^{\mu_1} \bar{\sigma}^{\mu_3}|2 \rangle [3|\bar{\sigma}^{\mu_2} \sigma^{\mu_4}|4] I_{2,1}^{\mu_1\mu_2\mu_3\mu_4} \Rightarrow 
        \frac{u^2}{9\epsilon^2}\langle 12 \rangle [34],\\
        &8)\, -\langle 1|\sigma^{\mu_1} \bar{\sigma}^{\mu_4}|4] [3|\bar{\sigma}^{\mu_2} \sigma^{\mu_3}|2\rangle I_{2,1}^{\mu_1\mu_2\mu_3\mu_4} \Rightarrow 
        0.
    \end{aligned}
\end{equation*}
Altogether one has
\begin{equation}
uDBub=\frac{1}{4} \cdot 4\frac{u^2}{9\epsilon^2} \langle 12 \rangle [34]=\frac{u^2}{9\epsilon^2} \langle 12 \rangle [34].
\end{equation}

For the t-channel we have the same configuration of fermion currents, resulting in the same expression with replacement $u\to t$
\begin{equation}
    tDBub=\frac{t^2}{9\epsilon^2} \langle 12 \rangle [34].
\end{equation}

\begin{figure}[ht]
\center{\includegraphics[scale=0.30]{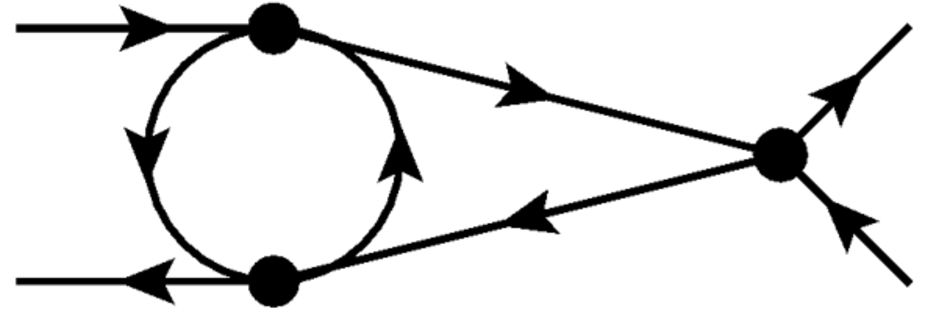}}
\caption{One of the fermion current configurations in the glass type topology}
\label{Glass}
\end{figure}

For the other kind of topology named the "glass" shown in Fig.\ref{Glass}

we provide the answers for the sum of different configurations of fermion currents in s-, t- and u-channels, respectively
\begin{equation}
sGlass=\frac{s^2}{12\epsilon^2} \langle 12 \rangle [34], \quad tGlass=\frac{7t^2}{144\epsilon^2} \langle 12 \rangle [34], \quad uGlass=\frac{7u^2}{144\epsilon^2} \langle 12 \rangle [34].      
\end{equation}

Then, for the two-loop $\langle 12 \rangle [34]$ amplitude we have 
\begin{equation}
\begin{aligned}
&RS_2=sDBub + 2\,sGlass = \frac{5s^2}{12\epsilon^2}\langle 12 \rangle [34],\\
&RT_2=tDBub + 2\,tGlass = \frac{5t^2}{24\epsilon^2}\langle 12 \rangle [34],\\ 
&RU_2=uDBub + 2\,uGlass = \frac{5u^2}{24\epsilon^2}\langle 12 \rangle [34],   
\end{aligned}
\end{equation}
\begin{equation}
     R_2(s,t) = RS_2+RT_2+RU_2 =\frac{5}{12\epsilon^2}(s^2+\frac{t^2}{2}+\frac{(-s-t)^2}{2}) \langle 12 \rangle [34].
              \label{R2}
\end{equation}

Using the interchanging rule, mentioned in the previous section, we can immediately get the result for the $\langle 14 \rangle [32]$ part
\begin{equation}
     L_2(t,s) =-\frac{5}{12\epsilon^2}(t^2+\frac{s^2}{2}+\frac{(-s-t)^2}{2}) \langle 14 \rangle [32].
\end{equation}

The correctness of the resulting expression can be verified by calculating of the corresponding diagrams. Finally, the full answer for the two-loop amplitude is as follows:
\begin{equation}
\begin{aligned}
A_4^{(2)} &= R_2(s,t) + L_2(t,s) \\
        &= \frac{5}{12\epsilon^2}(s^2+\frac{t^2}{2}+\frac{(-s-t)^2}{2}) \langle 12 \rangle [34] - \frac{5}{12\epsilon^2}(t^2+\frac{s^2}{2}+\frac{(-s-t)^2}{2}) \langle 14 \rangle [32].
\end{aligned}
\end{equation}

Below we show that this result can be obtained from the recurrence procedure, which can be created using the structure of the R - operation. This allows one to obtain the leading poles in any loop without a routine process of diagram calculation.

\subsection{Three loops}

\begin{figure}[ht]
\center{\includegraphics[scale=0.7, trim=0 0 0 0]{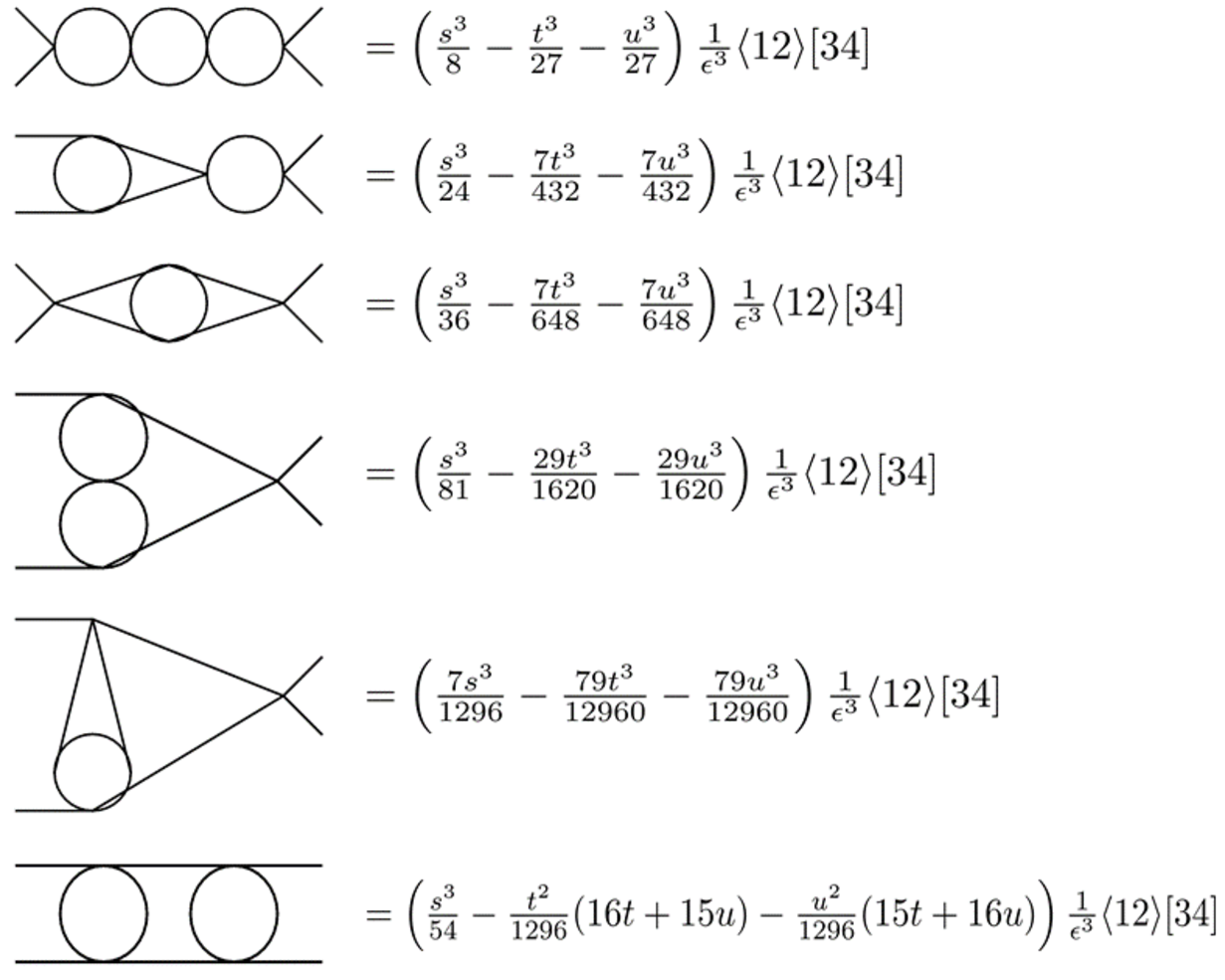}}
\caption{Results for three-loop diagrams}
\label{ThreeLoop}
\end{figure}

In this section we present the results for the calculation of the three-loop diagrams.  Each diagram configuration has 16 terms but many of them sum up to zero, so we present in Fig.\ref{ThreeLoop} only the final result for each kind of topology  without giving the details of the calculation  for s-, t- and u- channels, respectively.

Summing up all these answers allows us to write the answer for the amplitude in three loops
\begin{equation}
     R_3(s,t) = \frac{s}{432\epsilon^3}(196s^2+193st+193t^2) \langle 12 \rangle [34],
     \label{R3}
\end{equation}
\begin{equation}
     L_3(t,s) = -\frac{t}{432\epsilon^3}(196t^2+193st+193s^2) \langle 14 \rangle [32],
\end{equation}
\begin{equation}
\begin{aligned}
A_4^{(3)} &= R_3(s,t) + L_3(t,s) = \frac{s}{432\epsilon^3}(196s^2+193st+193t^2) \langle 12 \rangle [34] -\\&- \frac{t}{432\epsilon^3}(196t^2+193st+193s^2) \langle 14 \rangle [32].
\end{aligned}
\end{equation}

\newpage
\section{Summary}

Summarizing the results for the one-, two- and three-loop calculations we have:
\begin{equation}
A_4^{(1)} = \frac{5s}{6\epsilon} \langle 12 \rangle [34] - \frac{5t}{6\epsilon} \langle 14 \rangle [32],
\label{A1}
\end{equation}
\begin{equation}
A_4^{(2)} = \frac{5}{12\epsilon^2}(s^2+\frac{t^2}{2}+\frac{(-s-t)^2}{2}) \langle 12 \rangle [34] - \frac{5}{12\epsilon^2}(t^2+\frac{s^2}{2}+\frac{(-s-t)^2}{2}) \langle 14 \rangle [32],
\label{A2}
\end{equation}
\begin{equation}
    \begin{aligned}
       A_4^{(3)} = \frac{s}{432\epsilon^3}(196s^2+193st+193t^2) \langle 12 \rangle [34] - \frac{t}{432\epsilon^3}(196t^2+193st+193s^2) \langle 14 \rangle [32].
        \label{A3}
    \end{aligned}
\end{equation}

One can see that we have only two kinds of structures in each amplitude, namely, $\langle 12 \rangle [34]$ and $\langle 14 \rangle [32]$. This is a consequence of the usage of  two-dimensional Weyl spinors along with the spinor-helicity formalism in the calculation of the amplitude. Thus, we deal with these two structures, which appear independently in the corresponding Feynman diagrams but are mixed in the final answer for the whole amplitude. Due to that, we cannot factorize the tree level amplitude, but it is possible to divide the amplitude into two parts proportional to $\langle 12 \rangle [34]$ or $\langle 14 \rangle [32]$, respectively, which is useful for the study of the amplitudes in all orders of perturbation theory. 

The above expressions can be used as a playground for the formalism that we developed earlier ~\cite{we1, we2}.  Namely, using the locality of the counterterms after the application of the $R'$-operation, one can write down the recurrence relations that connect the counterterms in subsequent  orders of perturbation theory. These recurrence relations allow one to
 reproduce the coefficients that stand at spinor structures in (\ref{A1} - \ref{A3}) starting from one - loop. 
 The derivation of these relations is not that straightforward and we leave it together with the evaluation of the corresponding RG equations for another publication. Here we just give the main idea of how they are obtained and write them down in order to check our calculations and to see how the whole procedure works.

The derivation of the recurrence relations is based on the concept of the $R'$-operation \cite{BSh, Kazakov}.  The $R'$-operation is the incomplete renormalization operation which subtracts all UV subdivergences in subgraphs and leaves the overall UV divergence of the diagram.  Schematically for the four-point diagram it can be represented as follows, where the dotted line denotes the counter term obtained by the action of the $R'$-operation on the corresponding subgraph\cite{scalar}.

\begin{figure}[ht]
\center{\includegraphics[scale=0.4]{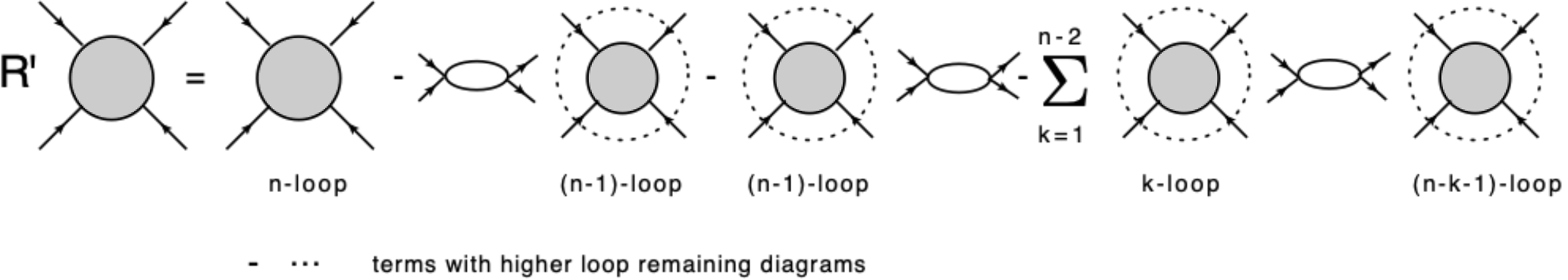}}
\vspace{0.4cm}
\caption{The $R'$-operation for the four-point diagram}
\label{R}
\end{figure}
\vspace{-0.5cm}

The key point here is that according to the Bogoliubov-Parasiuk theorem\cite{BP} the resulting expression after applying the $R'$-operation is local, i.e., does not contain non-local terms like $\frac{1}{\varepsilon^m}\log^k\frac{s}{\mu^2}$ for all $k$ and $m$. All the pole terms must be only polynomials of the Mandelstam variables. Requirement of cancellation of non-local terms  leads to relation between the pole terms the most significant of which connects the highest divergence in $n$-loops to the one-loop one. This allows one to get the relation between the $n$-loop and $(n-1)$-loop divergence. It is the recurrence relation we are talking about\cite{we1,we2}.

For the amplitudes of interest the advocated formalism allows one to get the  recurrence relations in the following form separately for both parts of the amplitude $\langle 12 \rangle [34]$ and $\langle 14 \rangle [32]$, where we write down the coefficients of these amplitudes omitting the obvious factors of $1/\epsilon^n$
\begin{equation}
\begin{aligned}
            nRS_n(s,u) = &-u\!\int_{0}^{1}\!\!\!dx \sum_{k=0}^{n-1}\sum_{p=0}^{k}\frac{[s(-u-s)]^p[x(1-x)]^{p+1}}{p!(p+1)!(p+1)^{-1}}\frac{d^p R_k(s,u')}{du'^p}\frac{d^p R_{n-1-k}(s,u')}{du'^p}
            \\
            &   -u\!\int_{0}^{1}\!\!\!dx \sum_{k=0}^{n-1}\sum_{p=0}^{k}\frac{[s(-u-s)]^p[x(1-x)]^{p+1}u'}{p!(p+1)!}\frac{d}{du'}\left[\frac{d^p R_k(s,u')}{du'^p}\frac{d^p R_{n-1-k}(s,u')}{du'^p}\right]
\\
            &+ s\!\int_{0}^{1}\!\!\!dx \sum_{k=0}^{n-1}\sum_{p=0}^{k}\frac{[s(-u-s)]^p[x(1-x)]^{p+1}}{p!(p+1)!}\left[\frac{d^p R_k(s,u')}{du'^p}\frac{d^p R_{n-1-k}(s,u')}{du'^p}\right]
\end{aligned}
    \label{RSUn}
\end{equation}
\begin{equation}
\begin{aligned}
            nRU_n(s,u) =&-u\!\int_{0}^{1}\!\!\!dx \sum_{k=0}^{n-1}\sum_{p=0}^{k}\frac{[u(-s-u)]^p[x(1-x)]^{p+1}}{p!(p+1)!(p+2)^{-1}}\frac{d^p R_k(s',u)}{ds'^p}\frac{d^p R_{n-1-k}(s',u)}{ds'^p}
            \\
            &   -u\!\int_{0}^{1}\!\!\!dx \sum_{k=0}^{n-1}\sum_{p=0}^{k}\frac{[u(-s-u)]^p[x(1-x)]^{p+1}s'}{p!(p+1)!}\frac{d}{ds'}\left[\frac{d^p R_k(s',u)}{ds'^p}\frac{d^p R_{n-1-k}(s',u)}{ds'^p}\right]
\end{aligned}
    \label{RUn}
\end{equation}
\begin{equation}
\begin{aligned}
   nRT_n(s,t) = nRU_n(s,u \rightarrow t )   \hspace{30em}
\end{aligned}
    \label{RTn}
\end{equation}
with $u'\rightarrow -sx$ in $RS_n$, $s'\rightarrow -ux$ in $RU_n$ and $s'\rightarrow -tx$ in $RT_n$. $R_k$ denotes the sum of the following contributions.
\begin{equation}
   R_k = RS_k(s,u)+RS_k(s,t)+RU_k(s,u)+RT_k(s,t)  
    \label{Rk}
\end{equation}
Similar relations can be written for $LS_n$, $LT_n$ and $LU_n$,  by interchanging $p_2 \leftrightarrow p_4$ and adding the anticommutation sign.

Using recurrence relations (\ref{RSUn}-\ref{RTn}) one can reproduce the one - loop results (\ref{R1}) by substituting the tree level expression (\ref{tree}) into these relations. We can also obtain two - loop results (\ref{R2}) using (\ref{R1}) and three - loop results (\ref{R3}) using (\ref{R2}) and (\ref{R1}) with the help of the same procedure. This way we can obtain the leading poles of the amplitude in an arbitrary number of loops, avoiding the diagram calculation, which is sometimes troublesome when dealing with fermions.

\section{V-A interaction}

In this section we briefly consider more physically interesting case of the four-fermion interaction, namely the V-A one with ${\cal O}=\gamma^\mu(1-\gamma^5)/2$ and provide the one-loop calculation of the scattering amplitude. 
%This interaction arises in describing the beta decay (see for example \cite{Schwartz}). 
For the V-A case the interaction Lagrangian has the form
\begin{equation}
{\cal L}_{int}= - \frac{G_F}{4} \bar \Psi \frac{\gamma^\mu(1 -\gamma^5)}{2} \Psi \bar \Psi \frac{\gamma^\nu(1 -  \gamma^5)}{2} \Psi = - \frac{G_F}{4} \bar \Psi \gamma^\mu P_L \Psi \bar \Psi \gamma^\nu P_L \Psi
\end{equation}

Following (\ref{vertex}) for the tree-level diagrams we have
\begin{equation}
\begin{aligned}
&\frac{1}{4}(\bar{\upsilon}_L(p_1) \gamma^\mu  u_L(p_2)\, g_{\mu\nu}\, \bar{\upsilon}_L(p_3) \gamma^\nu u_L(p_4) - \bar{\upsilon}_L(p_1) \gamma^\mu  u_L(p_4)\, g_{\mu\nu}\, \bar{\upsilon}_L(p_3) \gamma^\nu u_L(p_2))=
\\
&\frac{1}{4}(\langle 1|\gamma^\mu|2] \langle 3|\gamma_\mu|4] - \langle 1|\gamma^\mu|4] \langle 3|\gamma_\mu|2]) =\frac{1}{4}( 2\langle 13 \rangle [42]-2\langle 13 \rangle [24]) =  \langle 13 \rangle [42].   
\end{aligned}  
\end{equation}
So, at the tree-level one has
\begin{equation}
    A_4^{(0)} =  \langle 13 \rangle [42].
\end{equation}
Thus, we are left with only one structure. Repeating the one-loop calculations the same way as in Sec.4.1   we obtain in s-, t- and u- channels, respectively 
\begin{equation}
    S_1 = -\frac{4}{3}\frac{s}{\epsilon}\langle 13 \rangle [42], \qquad T_1 = -\frac{4}{3}\frac{t}{\epsilon}\langle 13 \rangle [42], \qquad U_1 = 4\frac{u}{\epsilon}\langle 13 \rangle [42].
\end{equation}

Finally, the one-loop amplitude is equal to 
\begin{equation}
    A_4^{(1)} = S_1+T_1+U_1 = \left(-\frac{16}{3}\frac{s}{\epsilon} -\frac{16}{3}\frac{t}{\epsilon}\right)\langle 13 \rangle [42].
\end{equation}
Factorizing the tree-level, one has
\begin{equation}
\frac{A_4^{(1)}}{A_4^{(0)}} = -\frac{16}{3}\frac{s+t}{\epsilon}.
\end{equation}

For the V-A case one can write down the same type of recurrence relations, which reproduce the leading singularities in all orders of perturbation theory
\begin{equation}
        nS_n(s,u) = -4s\!\int_{0}^{1}\!\!\! dx \sum_{k=0}^{n-1}\sum_{p=0}^{k}\frac{[s(-u-s)]^p[x(1-x)]^{p+1}}{p!(p+1)!(p+2)^{-1}}\frac{d^p A_4^{(k)}(s,u')}{du'^p}\frac{d^p A_4^{(n-1-k)}(s,u')}{du'^p}
    \label{Sn}
\end{equation}
\begin{equation}
         nT_n(t,u) = nS_n(s\rightarrow t,u)  \hspace{30em}
    \label{Tn}
\end{equation}
\begin{equation}
         nU_n(s,u) = 4u\!\int_{0}^{1}\!\!\! dx \sum_{k=0}^{n-1}\sum_{p=0}^{k}\frac{[u(-s-u)]^p[x(1-x)]^{p+1}}{p!(p+1)!(p+3)^{-1}}\frac{d^p A_4^{(k)}(s',u)}{ds'^p}\frac{d^p A_4^{(n-1-k)}(s',u)}{ds'^p}
    \label{Un}
\end{equation}
with $u'\rightarrow -sx$ in $S_n$, $u'\rightarrow -tx$ in $T_n$ and  $s'\rightarrow -ux$ in $U_n$. The function $A_4^{(k)}$ is the following  
\begin{equation}
   A_4^{(k)} = S_k(s,u)+T_k(t,u)+U_k(s,u)+U_k(t,u) 
    \label{Ak}
\end{equation}
We leave derivation and analysis of these relations for the other paper.
\section{Conclusions}

In this work, we have demonstrated that one can create all necessary topologies using the prescription of \cite{Paraskevas}. This also allows one to write down all the numerators, symmetry coefficients and other building blocks for the amplitude in any order of loop. However, the process of calculation in higher orders is still complicated, because the number of different types of diagrams depending on the configuration of the fermion flow grows as $2^{L+1}$ where $L$ is the number of loops.

We have seen that using the two-component spinor formalism, one can make calculations much easier compared to four dimensional Dirac spinors. This can also be used  for creating a computer algebra system package to calculate spinor chains. 

We resume that we have obtained the leading divergences up to three loops for the four-point $f f \rightarrow f f$ scattering amplitude for  the four fermion interaction model in four dimensions. We show that the number of independent structures in the amplitude in the scalar case can be reduced to two and the final answer in any loop is a linear combination of these two structures. As  for the V-A case, one has only one independent structure. We have checked the validity of our calculations confronting them with the recurrence relations that connect the subsequent orders of perturbation theory. 

Further analysis of the obtained results, recurrence relations and RG equations is in progress.

\bibliographystyle{hunsrt.bst}
\bibliography{bibliography}

\end{document}